\documentclass[conference]{IEEEtran}
\pdfoutput=1
\usepackage[hyphens]{url}
\usepackage{hyperref}
\hypersetup{breaklinks=true, colorlinks,allcolors=blue}
\usepackage[backend=biber, style=ieee]{biblatex}

\IEEEoverridecommandlockouts
\usepackage{amsmath,amssymb,amsfonts}
\usepackage{algorithmic}
\usepackage{graphicx}
\usepackage{textcomp}
\usepackage{xcolor}  
\usepackage{balance}
\usepackage{mathtools}  
\usepackage{kantlipsum} 
\usepackage{balance}
\usepackage{makecell}
\usepackage{orcidlink}

\usepackage{caption} 
\def\BibTeX{{\rm B\kern-.05em{\sc i\kern-.025em b}\kern-.08em
    T\kern-.1667em\lower.7ex\hbox{E}\kern-.125emX}}

\begin{document}

\title{Safeguarding Virtual Healthcare: A Novel Attacker-\\Centric Model for Data Security and Privacy}

\author{\IEEEauthorblockN{Suvineetha Herath\,\orcidlink{0009-0002-1262-7823}, Haywood Gelman\,\orcidlink{0009-0009-7208-1624}, John Hastings\,\orcidlink{0000-0003-0871-3622}, Yong Wang\,\orcidlink{0000-0002-1962-4847}}
\IEEEauthorblockA{\textit{The Beacom College of Computer and Cyber Sciences} \\
\textit{Dakota State University}\\
\{suvineetha.herath,haywood.gelman\}@trojans.dsu.edu, \{john.hastings,yong.wang\}@dsu.edu}
}

\maketitle

\begin{abstract}
  The rapid growth of remote healthcare delivery has introduced significant security and privacy risks to protected health information (PHI). Analysis of a comprehensive healthcare security breach dataset covering 2009-2023 reveals their significant prevalence and impact. This study investigates the root causes of such security incidents and introduces the Attacker-Centric Approach (ACA), a novel threat model tailored to protect PHI. ACA addresses limitations in existing threat models and regulatory frameworks by adopting a holistic attacker-focused perspective, examining threats from the viewpoint of cyber adversaries, their motivations, tactics, and potential attack vectors. Leveraging established risk management frameworks, ACA provides a multi-layered approach to threat identification, risk assessment, and proactive mitigation strategies. A comprehensive threat library classifies physical, third-party, external, and internal threats. ACA's iterative nature and feedback mechanisms enable continuous adaptation to emerging threats, ensuring sustained effectiveness. ACA allows healthcare providers to proactively identify and mitigate vulnerabilities, fostering trust and supporting the secure adoption of virtual care technologies.
\end{abstract}

\begin{IEEEkeywords}
Virtual Healthcare Model, Protected Health information, Data security, Privacy, Attacker-Centric Approach
\end{IEEEkeywords}

\section{Introduction}

The Virtual Healthcare Model (VHM) is a delivery approach that replaces face-to-face visits with phone or video consultations and frequently captures patient data asynchronously through survey tools or remote monitoring devices~\cite{hutchings2020virtual}. The advancement of network technologies has empowered healthcare systems to support patient-centric virtual healthcare, allowing patients to actively participate in healthcare decisions~\cite{woodsideVirtual2014}. Virtual healthcare also enables healthcare professionals and patients to interact remotely with the aid of an array of technologies including Web 4, big data, Internet of Things (IoT), IoMT~\cite{hancock2022virtual}, and artificial intelligence (AI)~\cite{healthtech2024ai}. Virtual health saw a dramatic surge during the pandemic, maintaining a level in October 2021 that was over 1300\% higher than pre-pandemic figures~\cite{shubham2022next}. There are several benefits of adopting VHM~\cite{shubham2022next,fortney2011re}. However, due to its extensive interconnectivity, VHM is inherently susceptible to cybersecurity risk~\cite{HealthcarePublicHealth2021}.

The increasing dependency on virtual care has made it easier for cyber criminals to access protected health information (PHI) which is considered a precious product in “the data-sharing economy” \cite{bagad2021data}. Between 2009 and 2023, there were 5,887 healthcare data breaches reported~\cite{hipaaJournal2023}, with the cost of healthcare data breaches consistently rising over the past 13 years at an average cost to organizations of \$10.93M USD~\cite{ibmDataBreachCost2023}. The rise in healthcare-related attacks over the previous three years isr validated by~\cite{verizon2023dbir}.

The growing reliance on virtual healthcare and the increasing interconnectivity of its systems have introduced vulnerabilities that traditional threat models cannot handle. For example, STRIDE-based models target vulnerabilities that compromise key security properties, including confidentiality, integrity, and availability (CIA). However, they do not adequately address the privacy harms that arise from healthcare data breaches. 

This study investigates the root causes of security incidents in virtual healthcare environments and develops the novel Attacker-Centric Approach (ACA) threat model, designed to comprehensively safeguard PHI from evolving cyber threats. The following critical research questions guide the research:
\begin{itemize}
  \item \textbf{RQ1}: What are the major threat categories and attack vectors endangering security and privacy of PHI in VHMs?
  \item \textbf{RQ2}: What are the frequencies of different types of data breaches related to virtual healthcare delivery.
\item \textbf{RQ3}: How can a comprehensive threat model be developed to effectively ensure protection of PHI in virtual healthcare environments?
\end{itemize}

In the remainder of the paper, Section \ref{methodology} describes the methodology, followed in Section \ref{litreview} by the findings from the literature review. Section \ref{threatmodel} introduces the threat model, and Section \ref{practical} discusses its practical use. Section \ref{analysis} provides an analysis of the study, and Section \ref{conclusion} concludes the paper.

\section{Methodology}\label{methodology}
A mixed-methods study~\cite{creswell2018research} was conducted to provide a comprehensive understanding of virtual healthcare threats identified in scientific literature and actual breach incidents.

\subsection{Qualitative Phase}

A thorough literature review gathered materials from a variety of reputable sources, including government records, legal documents, and industry reports. Key databases--ACM Digital Library, IEEEXplore, and EBSCO--were used to identify relevant literature. Focused search terms such as ``virtual healthcare'', ``cybersecurity'', and ``PHI breaches'' (Table \ref{tab:searchterms}) were employed to guide the search, looking for both foundational and highly cited works.

Backward and forward citation chaining (or snowballing)~\cite{hu2011definition} was used to follow the citation trail in papers backward and forward in time in order to explore the most influential studies on virtual healthcare security. The types of foundational and recent works found via backward and forward chaining appear in Table \ref{tab:searchterms}, broken down by search term. 

This process selected a total of 63 studies with an emphasis on recent developments and regulatory considerations. The results were analyzed to identify key threat categories and attack vectors (RQ1). Additionally, the review examined regulatory frameworks to evaluate their effectiveness in mitigating threats and vulnerabilities (RQ3).

\begin{table}[ht]
\caption{Search Terms and Citation Model}
\label{tab:searchterms}
\centering
\scriptsize
\resizebox{\columnwidth}{!}{%
\begin{tabular}{|p{2.3cm}|p{2.7cm}|p{3.5cm}|}
\hline
\textbf{Search Terms}       & \textbf{Backward Search Focus}                                       & \textbf{Forward Search Focus}                                          \\ \hline
Virtual care                & Foundational studies on the definition and evaluation of virtual care & Recent developments in healthcare data breaches, security, and privacy \\ \hline
Health data breaches        & Regulatory responses to historic data breaches                       & Recent studies on PHI breach impacts and security improvements         \\ \hline
Data protection regulations & HIPAA and HITECH                                                    & Latest advancements in data protection laws                           \\ \hline
Security                    & Frameworks for Securing Virtual Healthcare Systems                   & Recent improvements in healthcare security technologies               \\ \hline
Privacy regulations         & Early privacy laws impacting healthcare (e.g., HIPAA)                & Evolving privacy laws in virtual healthcare                           \\ \hline
Privacy harm                & Foundational studies on privacy risks and their impact               & Recent research on consequences and mitigation of privacy harms        \\ \hline
\end{tabular}
} 
\end{table}

\subsection{Quantitative Phase}

The quantitative phase of the study analyzed a healthcare data breach dataset of 4,753 incidents from 2009 to 2023~\cite{OCRPortal} to determine the frequency and severity of real-world healthcare data breaches (RQ2).

\subsection{Mixed-methods Integration}
The findings of this mixed-methods approach facilitated the development of a novel threat model, the Attacker-Centric Approach (ACA) (Section \ref{threatmodel}), specifically tailored to address the unique challenges of protecting PHI in virtual healthcare settings (RQ3). The ACA model integrates the identified threat categories and attack vectors (from RQ1), accounts for the frequencies and severities of data breaches (from RQ2), and aims to overcome the limitations of existing threat models and regulatory frameworks (related to RQ3). 

\subsection{Ethical Considerations}
The use of the publicly accessible and de-identified breach dataset poses minimal ethical concerns regarding privacy or sensitive information. Furthermore, while this research investigates vulnerabilities in virtual healthcare systems, rather than increasing public concern regarding the security and privacy of PHI, this research intends to reassure stakeholders through positive steps to enhancing security measures, ultimately fostering trust and confidence in virtual healthcare systems.

\section{Literature Review}\label{litreview}

Virtual healthcare operates within an interconnected network including individuals, communication technologies, IoT devices, sensors, network infrastructure, and cloud servers~\cite{NISTSP1500-201}, spanning multiple healthcare aspects including accessibility, diagnostics, consultations, and patient monitoring~\cite{harrison2022safe}. These systems monitor and manage cyber-physical connections, representing a novel generation of highly networked embedded control systems~\cite{humayed2017cyber}.

VHM must facilitate the secure and private exchange of health information. The integration of health information exchange (HIE) enhances these protections by ensuring that health information is exchanged securely and in compliance with national and organizational standards~\cite{hernandez2018official}.

PHI encompasses any information related to an individual's health status, healthcare, or payment for healthcare~\cite{HIPAA1996}. As virtual care becomes increasingly integral to healthcare delivery, its expanded accessibility introduces heightened risks, attracting cybercriminals seeking unauthorized access~\cite{hardcastle2020virtual}. As a result, violations related to PHI have become increasingly prevalent~\cite{wairimu2022modelling}. This significantly raises concerns related to data security, privacy, and regulatory compliance, necessitating a well-defined threat model to anticipate and mitigate potential security risks in virtual healthcare environments.

\subsection{Threat Categories and Attack Vectors in VHMs (RQ1)}\label{ltrq1}

The literature highlights the emergence of blended attacks~\cite{HISAC2019BlendedThreats}, which combine physical and cyber-attack strategies, posing significant threats to the security of PHI and personally identifiable information within healthcare systems. Such attacks exploit physical access to launch subsequent cyber-attacks, making healthcare settings particularly vulnerable to these composite threats~\cite{moses2016physical}.

Despite the risks, current security measures often overlook the physical component of threats, focusing primarily on information security. There is a notable gap in addressing physical security within current information security measures~\cite{mavroeidis2018framework}. The existing literature calls for robust security protocols that can safeguard against both types of threats, emphasizing the importance of integrating technological solutions with organizational and human factors to enhance the overall security posture of healthcare services.

\subsection{Frequencies of Data Breaches in Virtual Healthcare (RQ2)}

The distribution of individuals affected by data breaches is depicted in Table \ref{tab:data_breaches_distribution} for various covered entity types. The analysis indicates that healthcare providers (39.93\%) and health plans (38.83\%) are the most affected entities by healthcare data breaches, emphasizing the importance of robust security measures to safeguard sensitive healthcare information in these sectors. Business associates also have a significant influence on 21.22\% of the affected individuals, while healthcare clearinghouses only account for 0.03\%.

\begin{table}[!htbp]
  \caption{Individuals Affected by Data Breaches Per Entity Type}
\label{tab:data_breaches_distribution}
\resizebox{\columnwidth}{!}{%
\begin{tabular}{lcc}
\textbf{Covered entity type} & \textbf{Total individuals affected} & \textbf{Percent of overall total}  \\ \hline
Healthcare provider & 128,119,527 & 39.93\% \\ \hline
Health plan & 124,591,430 & 38.83\% \\ \hline
Business associate & 68,094,158 & 21.22\% \\ \hline
Healthcare clearinghouse & 83,486 & 0.03\% \\ \hline
\multicolumn{3}{l}{\textit{Data derived from~\cite{OCRPortal}.}}
\end{tabular}%
}
\end{table}

\begin{figure}[!htbp]
    \centering
    \includegraphics[width=0.9\linewidth]{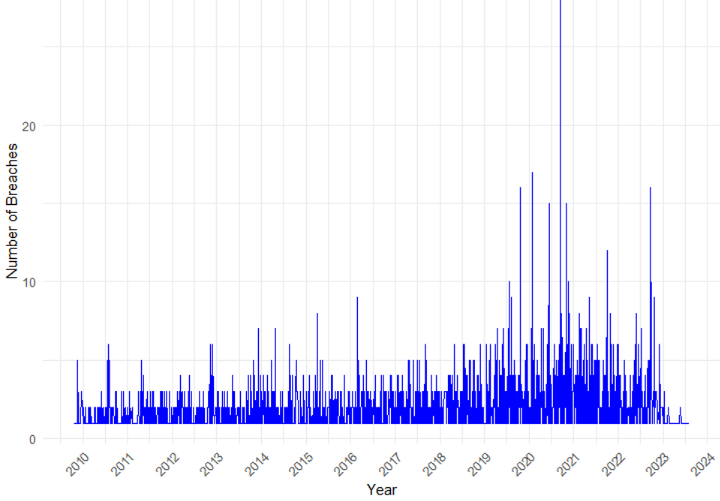}
    \caption{Frequency of Healthcare Data Breaches (2009–2023) \\ \textit{Data derived from~\cite{OCRPortal}.}}
    \label{fig:breach-type-individuals-affected}
\end{figure}

Fig. \ref{fig:breach-type-individuals-affected} shows the number of healthcare data breaches over time from the first reported breaches in the dataset until 2023. The frequency of healthcare data breaches has consistently trended upward over time, reflecting the growing challenges of securing sensitive health information in an increasingly digital healthcare environment. The graph indicates a significant increase in cyber-attacks during 2019-2021, which is likely due to the COVID-19 pandemic and the increased use of digital infrastructure, which makes it more vulnerable to breaches.

The dynamic nature of threats in healthcare systems stresses the need to establish a robust threat model that can enhance security measures and assist in compliance with regulatory frameworks. Comprehending the existing legal mandates in healthcare and advocating for pertinent legal adjustments is pivotal in fulfilling policy objectives, maintaining ethical standards, and propelling healthcare reform initiatives~\cite{harris2008contemporary}.

However, heath care privacy regulations are in continual flux, and managing the frequency of PHI access events within Electronic Health Records (EHR) systems presents operational challenges~\cite{MurphyHealthcareSecurityPrivacy}.  The United States is making a significant shift by migrating from HIPAA's privacy protections~\cite{HIPAA1996} to standard legal frameworks used by covered healthcare providers, especially in light of no federal data privacy law. The decision by the OCR to waive enforcement places the responsibility for managing security privacy directly on patients~\cite{bassan2020data}. Therefore, the evolving threat landscape in healthcare clearly highlights the necessity for a comprehensive threat model to ensure a resilient and compliant healthcare ecosystem.

\subsection{Existing Regulatory Frameworks and Threat Models (RQ3)}\label{frameworks}

\subsubsection{Regulatory Frameworks}

In the United States, the HIPAA and HITECH regulatory frameworks~\cite{HIPAA1996} play a crucial role in setting standards for the protection of PHI~\cite{hernandez2018official}. HIPAA ensures that health insurance coverage can be moved easily, reduces healthcare fraud, mandates standards for electronic health information, and stresses the protection of confidential patient data~\cite{hernandez2018official}. The HITECH Act~\cite{HITECHAct2009}, integrated with the 2009 American Recovery and Reinvestment Act, encourages the implementation of EHR to support patient data portability. Covered entities are identified in HIPAA as ``health plans, clearinghouses, and providers''~\cite{HHS2023CoveredEntities} engaging in electronic transmissions of health data for PHI transactions~\cite{HHS2023CoveredEntities}. Compliance with the HIPAA Security Rule is necessary for business associates, who manage PHI on behalf of covered entities~\cite{HHS2023CoveredEntities}. The rule mandates security of PHI by covered entities from known threats and unauthorized use or disclosure that could violate regulations~\cite{HHS2023CoveredEntities}. 

The \citetitle{congressional2023} report~\cite{congressional2023} addresses federal network security, critical infrastructure protection, data protection, and data privacy. However, ~\citeauthor{kosseff2017defining}~\cite{kosseff2017defining} highlights shortcomings in U.S. cybersecurity laws, noting the absence of a legal definition for cybersecurity. To effectively address network security and privacy concerns, it is necessary to seek amendments or comprehensive reforms to remedy technology limitations, prioritizing more robust provisions. 

HIPAA requires that healthcare providers and business associates mitigate cyber risk to data security and confidentiality of PHI. Despite robust regulations, significant challenges persist in fully securing healthcare data against the evolving landscape of cyber threats and rapidly advancing technology. Cyber-physical systems (CPS), a new generation of highly networked embedded control systems~\cite{humayed2017cyber} responsible for monitoring and controlling the physical realm, introduce significant security risks to both physical and digital domains due to their complex interdependencies among devices, networks, and actors. A fundamental challenge in CPS security arises from its components’ diverse and heterogeneous nature~\cite{humayed2017cyber}.

Continuous technological advances and increasing interconnectedness necessitate ongoing updates to  regulatory frameworks in order to maintain their effectiveness against new types of cyber threats. However, most regulations have been instituted reactively in response to past failures or incidents rather than proactively anticipating potential future issues~\cite{rasner2020third}.

To keep pace with advancing technology and the shifting tactics of cybercriminals, the healthcare industry must remain vigilant in updating security protocols and compliance measures. Furthermore, the continuously evolving nature of healthcare privacy research in response to technological and legal changes necessitates a well-structured and adaptive threat model. The model must consider all pertinent regulations and guide the strengthening of security postures in virtual healthcare systems by offering proactive measures to mitigate risks and enforce privacy protections.

\subsubsection{Threat Models}
To mitigate vulnerabilities, it is essential to identify and counteract threats from multiple perspectives, including authentication, authorization, privacy, adversary, and human-related risks~\cite{khatiwada2023threats}. Identifying and evaluating cyber threats using conventional risk assessment methods is difficult given the fast-changing nature of these threats~\cite{DiMase2015}. Threat modeling is an essential cybersecurity practice for enhancing security and privacy~\cite{shostack2014threat}.

The existing threat model, STRIDE, identifies major threat categories, including ``Spoofing, Tampering, Repudiation, Information Disclosure, Denial of Service, and Elevation of Privilege''~\cite{shostack2014threat}. However, STRIDE primarily focuses on security controls and does not fully address the unique challenges of virtual healthcare environments. For instance, mitigating emerging threat vectors, such as ransomware attacks targeting hospitals, demands more robust threat models beyond STRIDE. 

\begin{figure}[!htbp]
    \centering
    \includegraphics[width=0.9\linewidth]{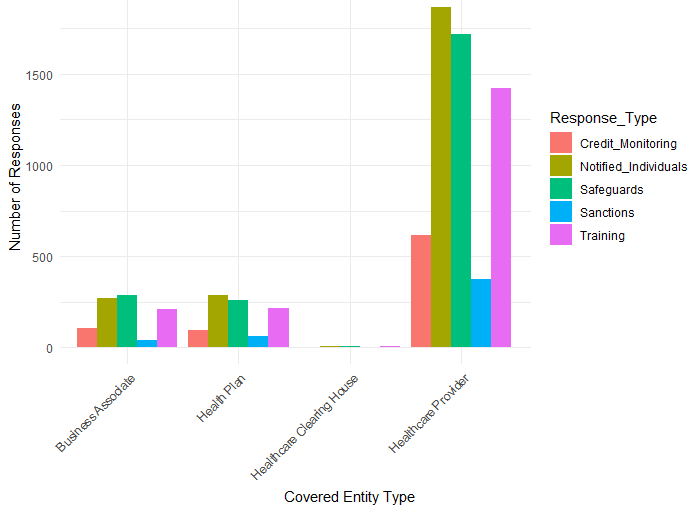}
    \caption{Number of Responses by Type per Covered Entity}
    \label{fig:response-types-entity}
\end{figure}

Fig. \ref{fig:response-types-entity} demonstrates that a significant portion of responses is focused on informing individuals after a breach, rather than actively implementing safeguards to prevent such breaches. This graph highlights that the emphasis remains on reactive measures, like notifying affected individuals, instead of adopting proactive security measures. Further, the figure shows response types by covered entity, where healthcare providers have the highest number of response actions, particularly in safeguards, notifying individuals, and training after breaches occur. 

\subsection{Threat Library}
The resulting threat library is based on the definition that a threat is ``any event or destruction, disclosure, modification of information, or denial of service''~\cite{NISTSP800-30r1}. Threat events can be organizationally determined by individual merit, in groups by commonality, or inferential in ordered succession~\cite{NISTSP800-30r1}. The data analysis (highlighted in Table \ref{tab:data_breaches_distribution} and Fig. \ref{fig:breach-type-individuals-affected})  identified key trends and vulnerabilities identified in historical breach incidents, and forms the foundation of the threat library (Table \ref{threat-library}) which provides a comprehensive overview of potential threats to healthcare systems. 

\begin{table}[htbp]
\renewcommand{\arraystretch}{1.2}
\caption{Threat Library based on Breach Analysis}
\label{threat-library}
\centering
\scriptsize
\resizebox{\columnwidth}{!}{%
\begin{tabular}{|p{1.0cm}|p{1.8cm}|p{4.3cm}|p{2.5cm}|}
\hline
\textbf{Category} & \textbf{Type of Threat} & \textbf{Description} & \textbf{Potential Threat Actors} \\
\hline
\textbf{Physical} & Natural Disasters & Events like floods, fires, etc., that disrupt healthcare systems & Natural conditions \\
\cline{2-4}
& Unauthorized Access & Physical breach allowing unauthorized entry to secure areas & Intruders, Trespassers \\
\cline{2-4}
& Theft or Vandalism & Theft of physical devices or data, or vandalism disrupting operations & Thieves, Vandalists \\
\hline
\textbf{Third Party} & Supply Chain Disruptions & Interruptions in third-party services affecting healthcare system operations & Business Associates, Supply Chain Issues \\
\cline{2-4}
& Utility Failures & Disruption of essential services like electricity or telecom & Utility Providers \\
\hline
\textbf{External} & Hacking/IT Incident & Cyberattacks targeting digital healthcare systems, such as ransomware & Cybercriminals, Hackers \\
\cline{2-4}
& Espionage & Covert efforts to gather sensitive healthcare information & State actors, Competitors \\
\hline
\textbf{Internal} & Unauthorized Access/Disclosure & Internal actors gaining unauthorized access to confidential healthcare information & Employees, Insiders \\
\cline{2-4}
& Insider Threats & Malicious or unintentional actions by employees affecting data security & Contractors, Employees \\
\hline
\end{tabular}
}
\end{table}

\section{The
  Attacker-Centric Approach (ACA) 
  Model}\label{threatmodel}

To bridge the identified gaps of existing frameworks in comprehensively addresseding evolving privacy and security, this study introduces a novel threat model, the Attacker-Centric Approach (ACA), specifically designed to enhance the security and privacy of VHMs. ACA adopts a holistic, attacker-focused perspective, examining potential threats from the viewpoint of threat actors, their motivations, tactics, and attack vectors. This perspective aids in identification of attack vectors common to all virtual care sectors, regardless of geographical boundaries or organization size.

\subsection{Development and Foundations}
The development of the ACA threat model was informed by
the findings from the previous section and established risk management frameworks and guideliness~\cite{NISTSP800-30r1,NISTSP800-61r2,cybersecurity2018framework}. 

\subsection{ACA Process}

\begin{figure}[!htbp]
    \centering
    \includegraphics[width=0.65\linewidth]{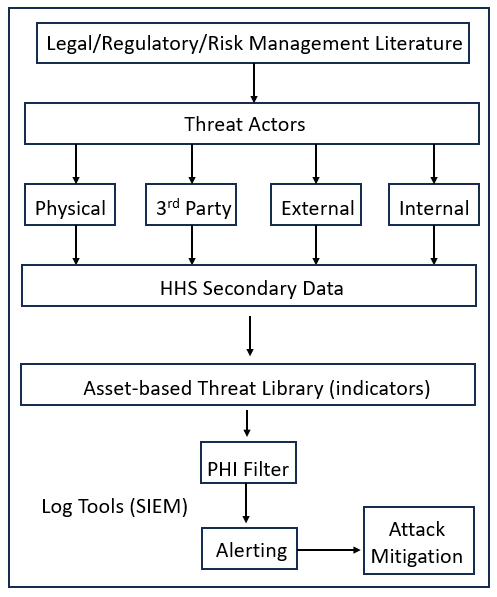}
    \caption{The Attacker-Centric Approach (ACA) Threat Model}
    \label{fig:enter-label}
\end{figure}

ACA aligns with \cite{NISTSP800-61r2}, which describes  a four-step process for
addressing computer security incidents: ``preparation, detection/analysis, containment/eradication/recovery, and post-incident activity''~\cite{NISTSP800-61r2}. The ACA flowchart, presented in Fig. \ref{fig:enter-label}, details the multiple layers of risk assessment and mitigation strategies that the model incorporates. 

ACA is a top-down processing model where a qualitative dataset is used to delineate threat actor types (input stage). A quantitative instrument is crafted from derived threat actor types which are then processed through the model to match with known attacks from HHS secondary cyber-attack data (processing stage). From this stage, the quantitative instrument is utilized to filter initial results through the threat library again to match for indicators of compromise using a PHI filter (filter stage). Integrated database results are then added to a Security Information and Event Management (SIEM) system for PHI-based alerting with attack mitigation rules measured by the reduction in attacks (output stage). Attack mitigation knowledge is then fed back into the model as new knowledge to improve ACA accuracy. 

The threat library (Table \ref{threat-library}) highlights major threats that can lead to healthcare security breaches. The library categorizes threat actors into four major categories: physical, third party, external, and internal. Physical threats are commonly caused by vulnerabilities in the physical environment~\cite{hutter2016}, where internal actors and their actions commonly cause insider threats~\cite{YuanWu2021}. In contrast, third-party risks are connected to vulnerabilities that arise from external entities, covering a range of potential risks and security breaches.

\subsection{Risk Assessment}

The model conducts a thorough analysis of the severity and potential impact of each identified threat. It utilizes historical data and patterns derived from network traffic, which are filtered through SIEM systems. This approach enables a detailed vulnerability analysis, identifying system weaknesses that could potentially be exploited. By identifying these vulnerabilities, the model effectively prioritizes mitigation strategies to strengthen system defenses against possible security breaches.

\subsection{Proactive Measures}
Based on the assessment, the model creates strategies to mitigate identified risks. This task may involve developing policy implementation and security protocols, enhancing existing defenses, conducting regular security audits, and supporting ongoing training programs to educate end users. The model also includes a feedback mechanism to update the threat library as new threats emerge or existing threats evolve, thus ensuring continuous relevance and effectiveness.

\section{Practical Application of ACA}\label{practical}

The following scenario is an illustration of the practical application of ACA with a large healthcare provider: 
\begin{enumerate}
\item \textit{Threat Identification}: The organization begins by utilizing ACA's threat library to identify relevant threat categories and specific threats applicable to their virtual healthcare platform. 

\item \textit{Risk Assessment}: Using ACA's risk assessment process, the organization analyzes the severity and potential impact of each identified threat. 

\item \textit{Proactive Mitigation}: Based on the risk assessment, the organization develops and implements proactive measures to mitigate identified risks. 

\item \textit{Continuous Improvement}: As new threats emerge or existing threats evolve, the organization leverages ACA's feedback mechanism to update the threat library and refine their mitigation strategies accordingly.
\end{enumerate}

By adopting ACA, the healthcare provider can significantly enhance the protection of PHI within their virtual healthcare platform. The attacker-centric approach provides a comprehensive understanding of potential threats, and the model's emphasis on continuous improvement ensures that the organization's security measures remain effective and relevant, promoting trust and confidence in their virtual healthcare services.

\section{Analysis of the ACA Threat Model}\label{analysis}

\subsection{Addressing Threat Categories and Attack Vectors (RQ1)}

Through a comprehensive threat library, ACA classifies potential threats into four major categories which encompass a range of specific threats, such as natural disasters, supply chain disruptions, social engineering attacks, and insider threats. Further, by adopting an attacker-centric perspective, ACA identifies potential attack vectors from the viewpoint of threat actors, their motivations, and their tactics. This approach ensures a thorough understanding of the diverse threat landscape, including physical, cyber, and blended attacks.

\subsection{Quantifying Frequencies of Data Breaches (RQ2)}

ACA incorporates the findings from the quantitative analysis of healthcare data breaches, which detailed the frequencies of different types of incidents, including ``hacking/IT incidents, improper disposal, loss, theft, and unauthorized access/disclosure''. By leveraging this comprehensive dataset, ACA accounts for the frequency of data breaches within the healthcare sector, underscoring the importance of addressing these specific threat categories within the ACA's threat library and risk assessment processes. The model's ability to prioritize mitigation strategies based on the potential impact of identified threats ensures that the most critical vulnerabilities are addressed promptly and effectively.

\subsection{Improving upon Existing Threat Models and Regulatory Frameworks (RQ3)}

ACA addresses the limitations of existing threat models and regulatory frameworks by adopting a holistic, attacker-centric approach that considers both physical and digital vulnerabilities. By examining potential threats from the perspective of attackers and their motivations, tactics, and potential attack vectors, ACA provides a more comprehensive understanding of the threat landscape. Furthermore, ACA's feedback mechanism incorporated into the model allows for the continuous incorporation of new knowledge and the adaptation of mitigation strategies to emerging threats. Future research could focus on developing detailed guidelines and best practices for implementing ACA within healthcare organizations, taking into account varying organizational structures, resources, and regulatory environments.

\section{Conclusion}\label{conclusion}

This study conducted a comprehensive investigation into the security and privacy challenges posed by VHMs, with a particular focus on the protection of PHI. The primary contribution of the work is the novel ACA threat model which effectively addresses identified threats and vulnerabilities to ensure the protection of PHI in VHM environments. ACA adopts an attacker-focused perspective, examining potential threats from the viewpoint of threat actors, their motivations, tactics, and attack vectors. This approach provides a comprehensive understanding of the threat landscape and enables proactive identification and mitigation of vulnerabilities.

ACA incorporates a comprehensive threat library, a robust risk assessment process, and proactive mitigation strategies. Notably, the model emphasizes continuous improvement through a feedback mechanism that allows for the iterative refinement of the threat library and the adaptation of mitigation strategies to emerging threats. This dynamic approach ensures the model's relevance and effectiveness in the ever-changing threat landscape of virtual healthcare. 

\balance
\printbibliography

\end{document}